\documentclass[%
 reprint,
 superscriptaddress,
 amsmath,amssymb,
 aps,
 showkeys
]{revtex4-2}

\usepackage[vietnamese,english]{babel}
\usepackage{graphicx}
\usepackage{dcolumn}
\usepackage{bm}
\usepackage[colorlinks=true,linkcolor=black,anchorcolor=black,citecolor=black,filecolor=black,menucolor=black,runcolor=black,urlcolor=black]{hyperref}

\begin{document}

\preprint{APS/123-QED}

\title{Inverse-designed dielectric cloaks for entanglement generation}

\author{A. Miguel-Torcal}
 \email{alberto.miguel@uam.es}
 \affiliation{Departamento de F\'isica Te\'orica de la Materia Condensada and Condensed Matter Physics Center (IFIMAC), Universidad Aut\'onoma de Madrid, E- 28049 Madrid, Spain.}
\author{J. Abad-Arredondo}
 \affiliation{Departamento de F\'isica Te\'orica de la Materia Condensada and Condensed Matter Physics Center (IFIMAC), Universidad Aut\'onoma de Madrid, E- 28049 Madrid, Spain.}%
\author{F. J. Garc\'ia-Vidal}
 \affiliation{Departamento de F\'isica Te\'orica de la Materia Condensada and Condensed Matter Physics Center (IFIMAC), Universidad Aut\'onoma de Madrid, E- 28049 Madrid, Spain.}
 \affiliation{Institute of High Performance Computing, Agency for Science, Technology, and Research (A*STAR), Connexis, 138632 Singapore.}
\author{A. I. Fern\'andez-Dom\'inguez}
 \email{a.fernandez-dominguez@uam.es}
 \affiliation{Departamento de F\'isica Te\'orica de la Materia Condensada and Condensed Matter Physics Center (IFIMAC), Universidad Aut\'onoma de Madrid, E- 28049 Madrid, Spain.}

\date{\today}

\begin{abstract}
We investigate the generation of entanglement between
two quantum emitters through the inverse-design engineering of
their photonic environment. By means of a topology-optimization
approach acting at the level of the electromagnetic Dyadic Green's
function, we generate dielectric cloaks operating at different
inter-emitter distances and incoherent pumping strengths. We show
that the structures obtained maximize the dissipative coupling
between the emitters under extremely different Purcell factor
conditions, and yield steady-state concurrence values much larger
than those attainable in free space. Finally, we benchmark our
design strategy by proving that the entanglement enabled by our
devices approaches the limit of maximum-entangled-mixed-states.
\end{abstract}

\maketitle


\section{Introduction}

The advent of quantum technologies relies on the design and
implementation of physical platforms able to support quantum
states involving a large number of elemental quantum systems
(qubits). Lately, the unprecedented control over light at the
sub-wavelength scale enabled by nanophotonics has emerged as a
promising resource for this purpose~\cite{OBrien2009,Uppu2021}.
Thus, different quantum functionalities exploiting the efficiency
and tunability of photon-assisted interactions in networks of
quantum emitters (QEs, such as atoms, molecules, quantum dots or
point defects in crystals) have been
proposed~\cite{Lodahl2017,Chang2018}. In this context, much
research attention has focused on entanglement formation in pairs
of qubits~\cite{Wootters1998}, a paradigmatic building block for
any quantum hardware, through their electromagnetic (EM) coupling
in different systems: one-dimensional optical
fibers~\cite{Zheng2013,Shahmoon2013,GonzalezTudela2013}, photonic
crystal cavities~\cite{Hughes2005,Samutpraphoot2020}, plasmonic
structures~\cite{GonzalezTudela2011,Hou2014,Dzsotjan2010},
metamaterials~\cite{Biehs2017,Jha2018} and chiral
waveguides~\cite{GonzalezBallestero2014,Pichler2015}. These
schemes found novel and feasible solutions to the long-standing
problem of bipartite entanglement
maximization~\cite{Ishizaka2000,Munro2001,Ficek2002} by making use
of the material and geometric toolsets accumulated over the years
of nanophotonics research.

Concurrently, the development of inverse design (ID) techniques
has made a strong impact in nanophotonics
research~\cite{Molesky2018,So2020,Wiecha2021}. ID algorithms have
proven to be very successful at enhancing, refining and optimizing
photonic functionalities~\cite{Piggott2015,Estakhri2019}. Among
the different members of the ID family, topology
optimization~\cite{Bendsoe2003} has contributed greatly to enlarge
the design space available for nanoscale optics~\cite{Jensen2011}.
Different implementations of this technique have revealed
unexpected and counterintuitive opportunities in areas as
different as optical circuitry~\cite{Dory2019}, second harmonic
generation~\cite{Lin2016opt}, nanoantennas~\cite{Christiansen2020}
and metasurfaces~\cite{Phan2019}. Only very recently, ID has been
transferred from the classical to the quantum regime, being
exploited to tailor nonclassical degrees of freedom of
nanophotonic fields. Thus, initial steps have shown the
manipulation of the local density of photonic
states~\cite{Lin2016prl,Mignuzzi2019} and the strength of
light-matter interactions~\cite{Bennett2020,Bennett2021}, as well
as the implementation of single photon
extractors~\cite{Chakravarthi2020} and the suppression of
inhomogeneous broadening effects in single-photon
transducers~\cite{Mishra2021}.

In this Article, we apply ID ideas to achieve photon-assisted
entanglement generation in QE pairs. In particular, we develop a
topology-optimization strategy to obtain dielectric cloaks for the
QEs that maximize the Wootters concurrence~\cite{Wootters1998} at
different inter-emitter distances. Both QEs are incoherently
pumped~\cite{Agarwal1990}, a technologically relevant
configuration~\cite{delValle2007,Weiler2012} that has been
overlooked in the recent literature on quantum nanophotonics.
After presenting our design method, we assess the concurrence
attained in the cloaks, revealing remarkable enhancements with
respect to free space. Next, we analyze the dielectric spatial
distribution within the devices, and offer insights into their
performance by investigating the character and strength of the QE
interactions as a function of the input parameters. Finally, we
benchmark the degree of entanglement in our ID structures against
those obtained using the
negativity~\cite{Vidal2002,Miranowicz2004} as the optimization
function, and show that our designs yield steady-state concurrence
values approaching the limit of
maximally-entangled-mixed-states~\cite{Ishizaka2000,Munro2001}.

\section{Physical System and Design Methodology}

The system under study consists of a pair of distant QEs, modelled
as identical two-level systems with perfect quantum yield, under
incoherent pumping. Assuming that they are only weakly coupled to
their dielectric environment, and after tracing out the EM degrees
of freedom, the master equation~\cite{Dzsotjan2010,Dung2002}
describing the photon-assisted interactions between them has the
form
\begin{equation}
\imath\Big[\rho,H\Big]+\sum_{i,j}\frac{\gamma_{ij}}{2}\mathcal{L}_{ij}(\rho)+\sum_{i}\frac{P_i}{2}\mathcal{L}^{\prime}_{ii}(\rho)=0,
\label{MEq}
\end{equation}
with $H=\sum_i\omega\sigma_i^{\dagger}\sigma_i+\sum_{i\neq j}
g_{ij}\sigma_i^{\dagger}\sigma_j$, and where the indices $i=1,2$
and $j=1,2$ label the two emitters. The first term in
(\ref{MEq}) accounts for the QE-QE coherent coupling, with
$\sigma_i$ ($\sigma^{\dagger}_i$) being the annihilation
(creation) operator for the emitter $i$. The second one includes
Lindblad superoperators of the form
$\mathcal{L}_{ij}(\rho)=2\sigma_j\rho\sigma_i^{\dagger}-\sigma_i^{\dagger}\sigma_j\rho-\rho\sigma_i^{\dagger}\sigma_j$
and reflects the dissipative interaction between the QEs ($i\neq
j$), as well as their radiative decay ($i=j$). Finally, the incoherent pumping of both QEs is expressed in terms of Lindblad superoperators
$\mathcal{L}^{\prime}_{ii}(\rho)=2\sigma^{\dagger}_i\rho\sigma_i-\sigma_i\sigma_i^{\dagger}\rho-\rho\sigma_i\sigma_i^{\dagger}$.

There are four different sets of parameters in (\ref{MEq}).
First, the QEs natural frequency, $\omega$, which we set to {3.1
eV} ($\lambda=400$ nm). Note that in this frequency range, metals
sustain highly confined surface plasmon modes, which have been
exploited recently in other nanophotonic proposals for
entanglement
generation~\cite{GonzalezTudela2011,Hou2014,Dzsotjan2010,MartinCano2011,Gangaraj2015}.
Secondly, the incoherent pumping rate on each QE, which is assumed
to be symmetric, $P_i=P_j=P$, and can be externally controlled by,
for instance, optical or electrical
means~\cite{delValle2007,Weiler2012}. Last, the coherent and
dissipative coupling strengths, that can be expressed as a
function of the Dyadic Green's function $\mathbf{G}(\mathbf{r},
\mathbf{r'},\omega)$~\cite{bookNovotny2012} for the dielectric
environment, evaluated at the QEs natural frequency. They read
$g_{i j}=\omega^{2}{\rm
Re}\{\mathbf{p}^*\mathbf{G}(\mathbf{r}_{i}, \mathbf{r}_{j},
\omega)\mathbf{p}\} /\hbar \varepsilon_{0} c^{2}$ and $\gamma_{i
j}=2\omega^{2}{\rm Im}\{\mathbf{p}^*\mathbf{G}(\mathbf{r}_{i},
\mathbf{r}_{j}, \omega)\mathbf{p}\}/\hbar \varepsilon_{0} c^{2}$,
respectively, where $\mathbf{p}$ is the transition dipole moment
of the QEs and $\mathbf{r}_{i,j}$, their position. In open
non-chiral EM systems~\cite{Downing2019}, the coupling constants
($i\neq j$) fulfil $\gamma_{ij}=\gamma_{ji}$ and $g_{ij}=g_{ji}$.
For $i=j$, $\gamma_{i i}=F(\omega,\mathbf{r}_{i})\gamma_0$ gives
the QE decay rate, where $F(\omega,\mathbf{r}_{i})$ is the Purcell
factor it experiences, and
$\gamma_0=\omega^{3}|\mathbf{p}|^2/3\pi\hbar \varepsilon_{0}
c^{3}$ its decay rate in free space.

With the density matrix, $\rho$, expressions for the expectation
values of any physical observable for the system (or in our case,
of an entanglement witness) can be constructed, which present an
explicit dependence on the master equation parameters and,
therefore, on the Dyadic Green's function. Taking a given physical
quantity as the target function, our ID approach seeks for the QEs
dielectric environment (the spatial distribution of the
permittivity around them) that optimizes (generally maximizes or
minimizes) it. We follow a topology-optimization-inspired
algorithm whose starting point is free space, i.e.,
$\epsilon_1(\mathbf{r})=1$ in the whole domain of interest. The
iterative procedure can be briefly described as follows: Each
iteration step, labelled as $n$, consists in a spatial sweep
around the QEs. At each position, $\mathbf{r}_k$ (of volume
$\delta V_k$), an small increment is introduced in the dielectric
constant,
$\epsilon'_{n+1}(\mathbf{r}_k)=\epsilon_n(\mathbf{r}_k)+\delta\epsilon$
(note that, for clarity, we have introduced index $k$ to reflect
the spatial discretization of the permittivity map). This modifies
the target function through the Dyadic Green's function. If this
local dielectric alteration contributes towards the optimization,
then $\epsilon_{n+1}(\mathbf{r}_k)=\epsilon'_{n+1}(\mathbf{r}_k)$.
Otherwise, the increment is discarded and
$\epsilon_{n+1}(\mathbf{r}_k)=\epsilon_n(\mathbf{r}_k)$.

In principle, the approach introduced above requires computing
$\mathbf{G}(\mathbf{r}_i, \mathbf{r}_j,\omega)$ for each local
dielectric increment $k$ and each iteration step $n$. This is, in
general, largely computationally demanding. However, for small
enough $\delta\epsilon$, the modification in the Dyadic Green's
function induced by the permittivity change at $\mathbf{r}_k$ can
be described perturbatively. Thus, keeping only the first term in
the Born scattering series~\cite{Bennett2020,bookNovotny2012}, we
have 
\begin{equation}
\delta'_k\mathbf{G}_{n+1}(\mathbf{r}_i,\mathbf{r}_j,\omega)=\tfrac{\omega^2}{c^2}\mathbf{G}_n(\mathbf{r}_i,\mathbf{r}_k,\omega)\delta\epsilon\mathbf{G}_n(\mathbf{r}_k,\mathbf{r}_j,\omega)\delta
V_k, \label{dG}
\end{equation}
whose effect in the target function still needs to be evaluated.
If this local variation of the permittivity contributes to its
optimization, $\delta\epsilon$ is kept and
$\delta_k\mathbf{G}_{n+1}(\mathbf{r}_i,\mathbf{r}_j,\omega)=\delta'_k\mathbf{G}_{n+1}(\mathbf{r}_i,\mathbf{r}_j,\omega)$,
while $\delta\epsilon$ is discarded and
$\delta_k\mathbf{G}_{n+1}(\mathbf{r}_i,\mathbf{r}_j,\omega)=0$
otherwise. As a result of the sweep in $k$ a new, complete,
permittivity map, $\epsilon_{n+1}(\mathbf{r})$, is obtained, for
which the Dyadic Green's function
$\mathbf{G}_{n+1}(\mathbf{r}_i,\mathbf{r}_j,\omega)$ can be
calculated through EM simulations. Moreover, the convergence of
the algorithm can be easily tested after each iteration step by
computing
\begin{equation}
\mathbf{G'}_{n+1}(\mathbf{r}_i,\mathbf{r}_j,\omega)=\mathbf{G}_{n}(\mathbf{r}_i,\mathbf{r}_j,\omega)+\sum_k\delta_k\mathbf{G}_{n+1}(\mathbf{r}_i,\mathbf{r}_j,\omega),
\label{Gn+1}
\end{equation}
and verifying that
$\mathbf{G'}_{n+1}(\mathbf{r}_i,\mathbf{r}_j,\omega)=\mathbf{G}_{n+1}(\mathbf{r}_i,\mathbf{r}_j,\omega)$
within the accuracy preset for the algorithm. Importantly, using
that
$\mathbf{G}_n(\mathbf{r}_k,\mathbf{r}_j,\omega)=\mathbf{G}^{\rm
T}_n(\mathbf{r}_j,\mathbf{r}_k,\omega)$, the evaluation of
(\ref{dG}) in all space only requires two EM simulations. For
the iteration $n+1$, these correspond to the spatial profile of
the electric fields radiated by both QEs, independently, within
the permittivity map $\epsilon_n(\mathbf{r})$.

\begin{figure}[h!]
\centering
\includegraphics[width=\linewidth]{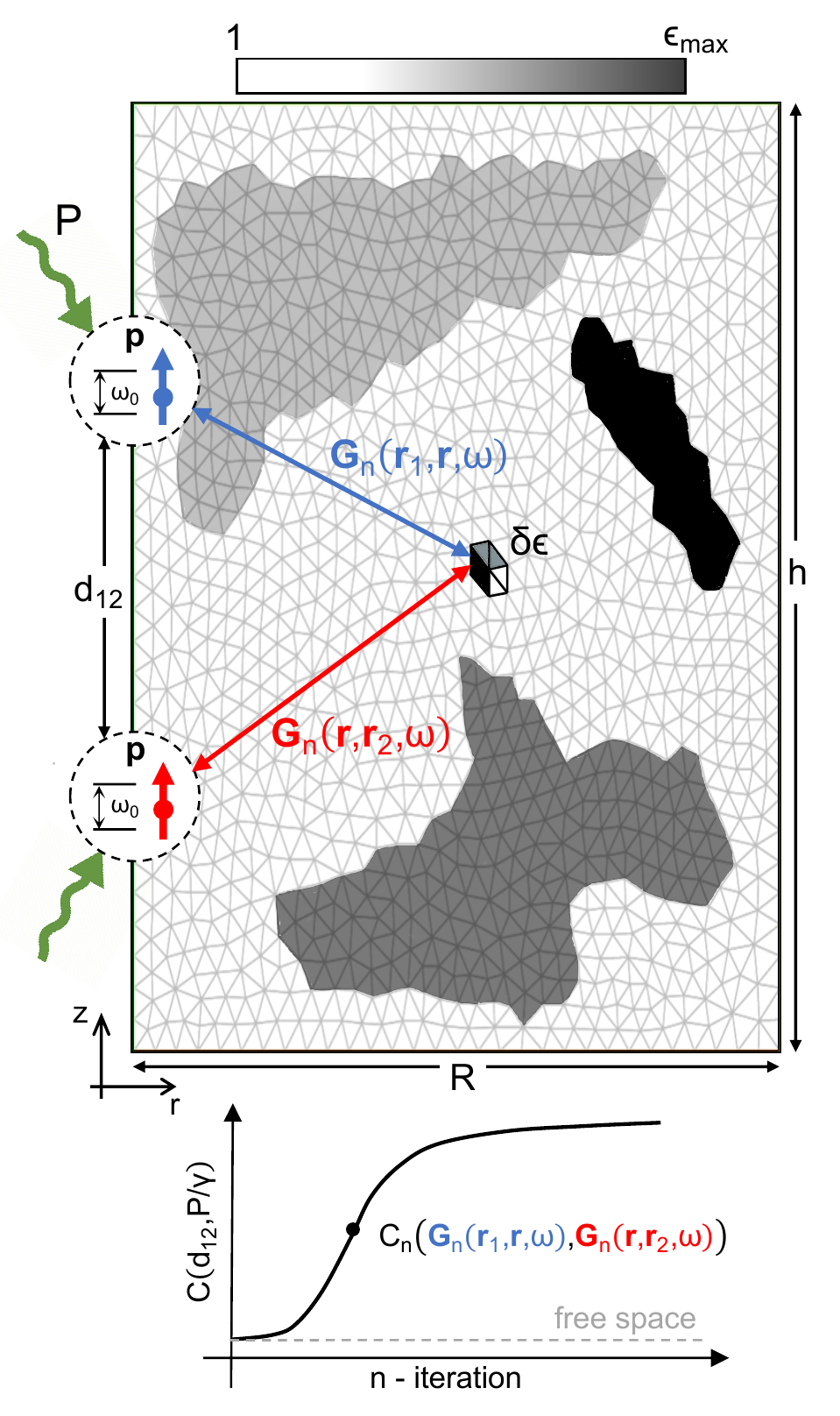}
\caption{(a) Sketch of the topology-optimization design of a
cylindrical cloak, of radius $R$ and height $h$, that maximizes
the Wootters concurrence between two QEs aligned along
$z$-direction and separated by a distance $d_{12}$. The light grey
mesh renders the spatial discretization, while the device
permittivity at iteration $n$ is coded from white ($\epsilon_n=1$)
to black ($\epsilon_n=\epsilon_{\rm max}$). The bottom panel
illustrates the concurrence, $C_n$ for the QE pair, as a function
of the iteration step, $n$.} \label{fig:1}
\end{figure}

\begin{figure}[t!]
\centering \vspace{0.1 cm}
\includegraphics[width=\linewidth]{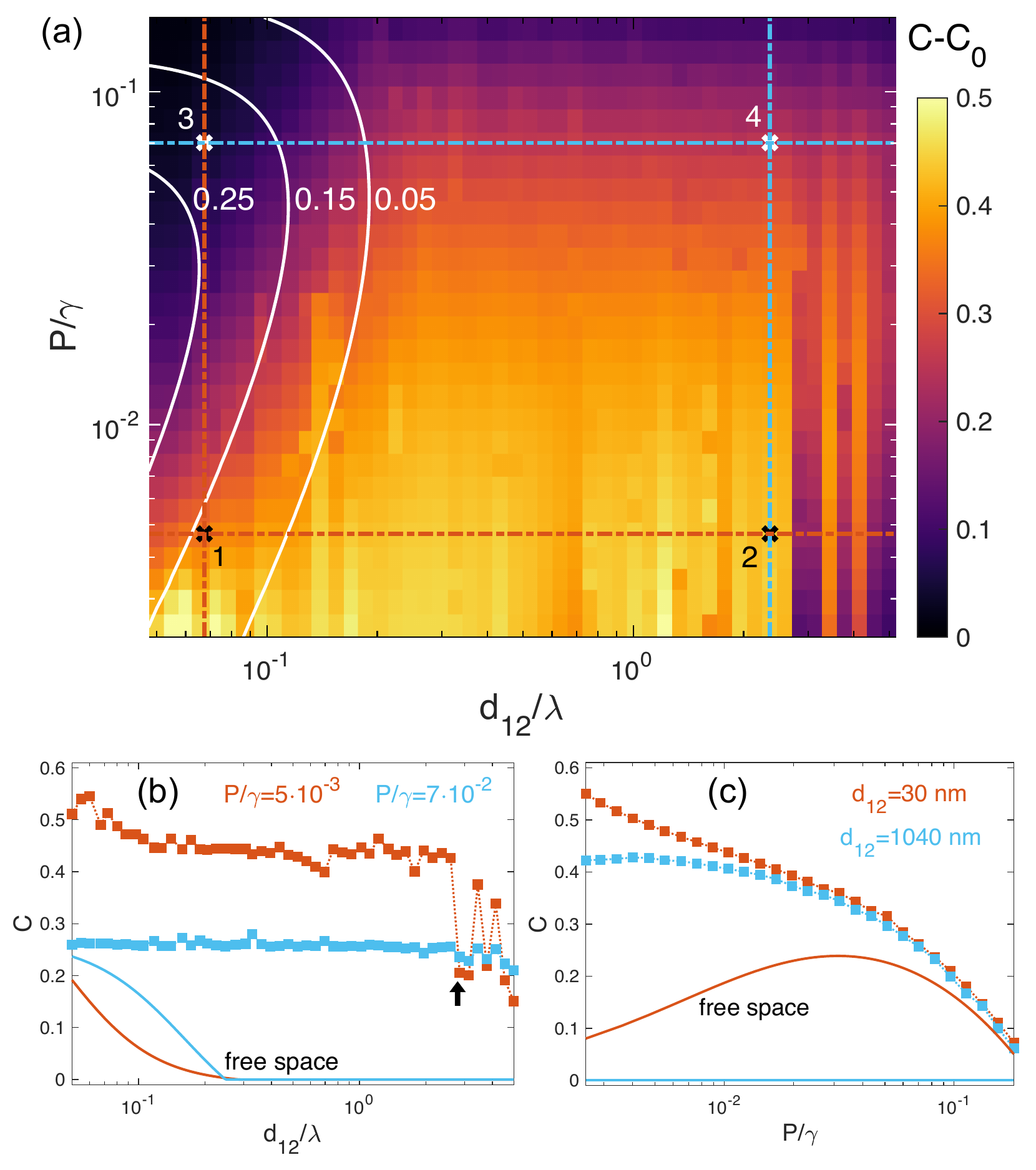}
\vspace{-0.5 cm} \caption{(a) Entanglement generation efficiency,
$C-C_0$, versus inter-emitter distance and pumping rate for
$50\times 27=1350$ ID cloaks. White solid lines correspond to
QE-QE systems yielding three different values of the free-space
concurrence, $C_0$. Vertical and horizontal lines indicate the
configurations considered in the panels below. (b) Cloak-induced
(square dots) and free-space (solid lines) concurrences versus
inter-emitter distance for low (orange) and high (blue) pumping.
(c) Same as (b) but versus pumping strength and for short (orange)
and long (blue) inter-emitter distance.} \label{fig:2}
\end{figure}

Figure \ref{fig:1} illustrates the ID approach described above. We
employ the finite-element solver of Maxwell's Equations
implemented in Comsol Multiphysics$^{\rm TM}$, whose spatial
discretization is represented by the light grey thin mesh. Note
that we employ this grid for the permittivity spatial distribution
as well. In our designs, both QEs are aligned, with their dipole
moments parallel to the axis that connects them ($z$-direction).
This way, we can exploit the azimuthal symmetry of the system to
solve the 3D EM problem within the $rz$-plane only. The size of
the cylindrical cloaks is given by the parameters $R$ and $h$,
while the distance between the QEs is $d_{12}$ (taken as an input
parameter). $\epsilon_{\rm max}$ is the maximum dielectric
constant in the device, which varies from one design to another.
In our calculations, we have set a threshold, $\epsilon_{\rm
max}\leq 9$, which corresponds to semiconductor materials such as
GaP~\cite{Cambiasso2017}. As anticipated, we take the Wootters
concurrence as a measure of entanglement and therefore, as the
optimization (in this case, maximization) function. This is
defined in terms of the eigenvalues of the matrix $\rho T\rho^*
T$, where $T$ is the anti-diagonal matrix with elements
$\{-1,1,1,-1\}$. For our system, we have
\begin{equation}
C=C(\rho)=2\,{\rm
max}\{0,|\rho_{12}|-\sqrt{\rho_{00}\rho_{33}}\},\label{C}
\end{equation}
where $\rho_{00}=\langle g_1g_2|\rho|g_1g_2 \rangle$ and
$\rho_{33}=\langle e_1e_2|\rho|e_1e_2 \rangle$ are the population
of the ground and biexciton states and $\rho_{12}=\langle
e_1g_2|\rho|g_1e_2 \rangle$ is the coherence between single
excitation states. A maximally entangled (completely untangled)
state is characterized by $C=1$ ($C=0$). The lower panel of
Figure \ref{fig:1} sketches the concurrence maximization, where
$C_n=C_n(\mathbf{G}_n(\mathbf{r}_1,\mathbf{r},\omega),\mathbf{G}_n(\mathbf{r},\mathbf{r}_2,\omega))$
corresponds to its value at iteration $n$ (note that we have made
explicit its dependence on the Dyadic Green's function connecting
the QE positions and the whole volume of the dielectric cloak).

\section{Results}

Figure \ref{fig:2} investigates the performance of the dielectric
cloaks (with dimensions $R=3.75\lambda$ and $h=10\lambda$)
obtained through the topology-optimization procedure described
above. The color map in Figure \ref{fig:2}(a) displays $C-C_0$, the
difference between the QE-QE concurrence, $C$, for 1350 ID
structures and their free-standing counterpart, $C_0$ (obtained
from the evaluation of $C$ for free-space master equation
parameters). This quantity, which we take as a measure of
entanglement generation efficiency, is rendered against the
inter-emitter distance, $d_{12}$ (normalized to the QE wavelength,
$\lambda=400$ nm), and pumping strength, $P$ (normalized to
$\gamma=\gamma_{11}=\gamma_{22}$, the emitter decay rate). In free
space, $\gamma=\gamma_0$, while $F(\omega,\mathbf{r}_{1,2})\neq1$
within the ID devices. Note that, although this is not a
constraint imposed in our design strategy, both QEs experience the
same Purcell factor,
$F(\omega,\mathbf{r}_{1})=F(\omega,\mathbf{r}_{2})$, in all the
structures generated. Thus, the horizontal axis in
Figure \ref{fig:2}(a) sets the minimum optical path between the QEs,
while the vertical one serves as a measure of their steady-state
population ($\rho_{11}^{\rm iso}=P/(\gamma+P)$ for the QEs in
isolation~\cite{Downing2020}). Both are in log scale, with a
logarithmic density of system configurations as well. The white
solid lines correspond to the pumping and distance conditions
yielding three different values of $C_0$: 0.25, 0.15 and 0.05. The
latter can be identified as the boundary beyond which the
free-space concurrence vanishes, as
$|\rho_{12}|<\rho_{00}\rho_{33}$ in $C$. Remarkably, it is
exactly in this region where the dielectric cloaks perform best,
leading to $C-C_0=C\approx0.5$ for distances up to
$2.85\lambda=1140$ nm and low pumping rate. We anticipate here
that this concurrence enhancement approaches the limit of
maximum-entangled-mixed-states~\cite{Ishizaka2000}. At smaller
$d_{12}$ and larger $P$, where $C_0$ is not negligible, their
efficiency worsens. This shows that, rather than enhancing $C$,
our ID devices are able to generate entanglement in QE-QE
configurations where the free-space concurrence vanishes. 

Figure \ref{fig:2}(b) analyzes the dependence of the entanglement
generation efficiency on the inter-emitter distance for two
different pumping rates. The square dots plot $C$ as a function of
$d_{12}/\lambda$ along the two horizontal lines indicated in panel
(a). For comparison, $C_0$ in the same pumping conditions are
plotted in solid lines. We can observe that the cloak-induced
concurrence presents little sensitivity to the QE-QE distance at
high pumping (blue), and it decays slowly with distance at low $P$
(orange). Both set of data present an abrupt reduction in $C$ at
$d_{12}=1140\,{\rm nm}$ ($2.85\lambda$, marked by a black vertical
arrow) followed by oscillations, more apparent at low pumping. As
shown below, these features originate from finite size effects,
which become stronger as the inter-emitter distance approaches the
cloak dimensions. In Figure \ref{fig:2}(c), the effect of the pumping
strength is explored. It plots $C$ and $C_0$ along the vertical
lines in panel (a). The former decays monotonically with
$P/\lambda$ in a very similar way for the two distances
considered. There exist differences at very small $P$, where the
cloaks for short inter-emitter distance (orange) yield larger $C$
than the ones for long distance (blue).

\begin{figure}[h!]
\centering
\includegraphics[width=\linewidth]{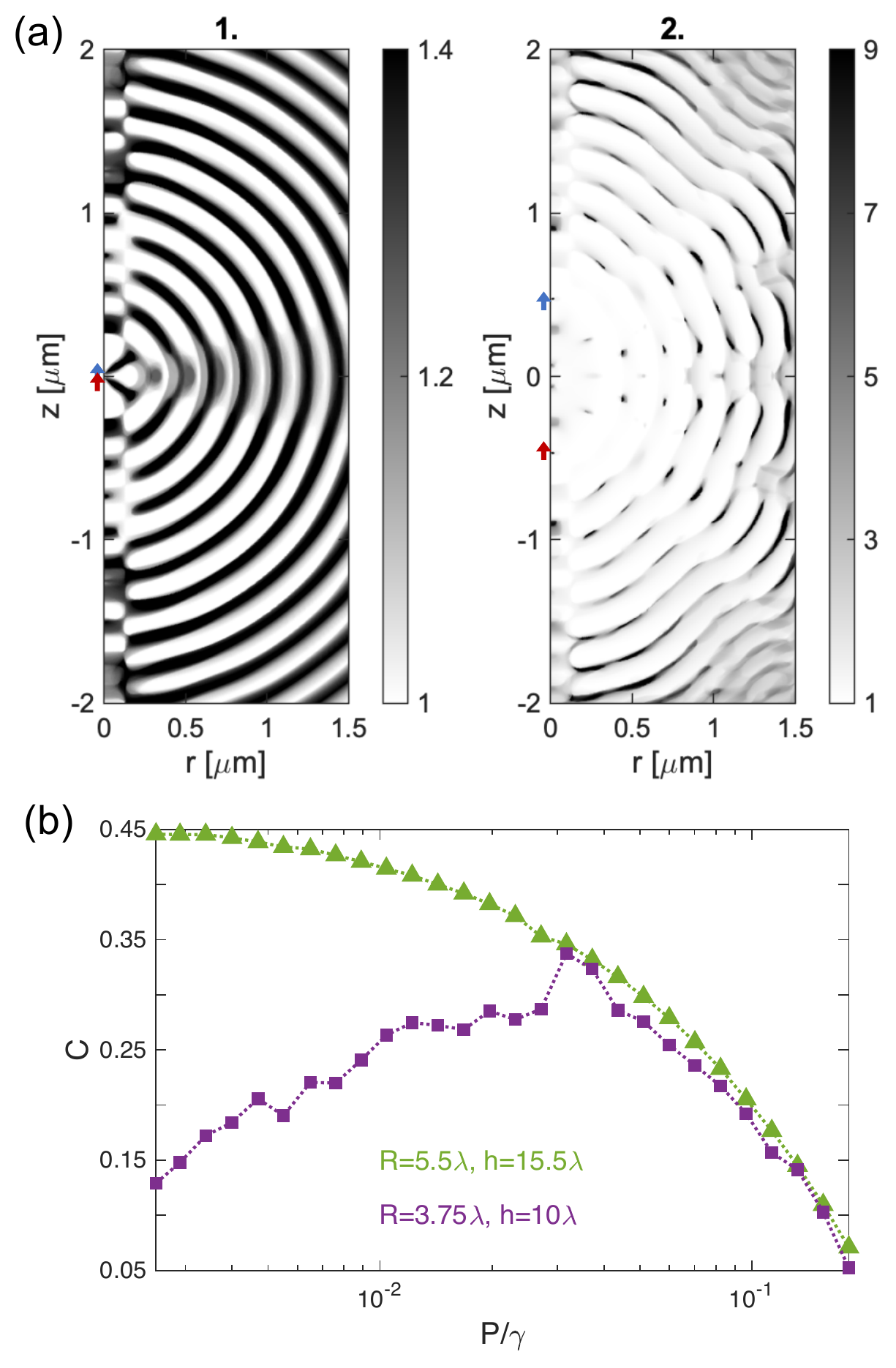}
\vspace{-0.5 cm}
 \caption{(a) Permittivity maps for the ID cloaks
labelled as 1 (left) and 2 (right) in Figure \ref{fig:2}(a). The
dielectric constant is represented by white-to-black linear scales
with different $\epsilon_{\rm max}$. (b) Concurrence versus
pumping rate for devices of two different sizes: $R=3.75\lambda$
and $h=10\lambda$ (purple) and $R=5.5\lambda$ and $h=15.5\lambda$
(green). Both set of cloaks operate at $d_{12}=2.85\lambda$, QE-QE
distance indicated by the vertical black arrow in
Figure \ref{fig:2}(b).} \label{fig:3}
\end{figure}

The dielectric distribution, $\epsilon(\mathbf{r})$, for the
cloaks labelled as 1 and 2 in Figure \ref{fig:2}(a) is shown in
Figure \ref{fig:3}(a). These are chosen in the low pumping regime
($P/\gamma=5\cdot 10^{-3}$), where $C-C_0$ is largest. Note that
exploiting the cylindrical symmetry of the designs, the
permittivity maps are fully characterized within the $rz$-plane,
with the QE positions indicated by red and blue arrows along the
$z$-axis. The grey scale codes the dielectric constant linearly
from 1 (white) to $\epsilon_{\rm max}$ (black). In the left panel
(1), the QEs are only a few nanometers apart ($d_{12}=30$ nm), and
the permittivity contrast is small, $\epsilon_{\rm max}=1.4$. We
can identify two different structures in the cloak. First, a
narrow waveguide along $z$-direction, with radius $\lambda/2=200$
nm, approximately, that mediates the QE-QE interactions. Around
it, a periodic and concentric pattern is apparent, with elements
that act as reflectors that reduce radiation leakage. The
dielectric distribution in this region is mainly binary,
$\epsilon(\mathbf{r})=1$ or $\epsilon_{\rm max}$, except around
the $z=0$ plane, along which dipole radiation power is maximum and
the permittivity acquires intermediate values. The right panel of
Figure \ref{fig:3}(a) corresponds to device 2, the QEs are farther
apart ($d_{12}=950$ nm) and the maximum permittivity is much
larger ($\epsilon_{\rm max}=9$). This is the threshold value set
for the topology-optimization algorithm, whose convergence
required significantly more iterations than in the left panel. The
resulting $\epsilon(\mathbf{r})$ still resembles device 1. The
dielectric contrast along $z$-axis is now much smaller than around
it. This is specially evident between the QEs. The geometry of the
reflecting elements is more complex, with much sharper and
isolated high-permittivity scatterers that overlap with multiple
periodic-like patterns of moderate dielectric constant. In
contrast to the left panel, the cloak is far from binary, with
$\epsilon(\mathbf{r})$ varying smoothly in some spatial regions
and much more abruptly in others. The underlying similarities
between devices 1 and 2 in Figure \ref{fig:3}(a) indicates that both
ID cloaks generate entanglement by simultaneously engineering the
mutual coupling between the QEs and minimizing their emission into
free-space.

Figure \ref{fig:3}(b) reveals the impact of the finite size of the ID
cloaks in their performance. It plots the concurrence versus
pumping strength for devices operating at
$d_{12}=2.85\lambda=1140\,{\rm nm}$, the distance indicated by a
black arrow in Figure \ref{fig:2}(b). The purple dots correspond to
the structures in that panel, with dimensions $R=3.75\lambda$ and
$h=10\lambda$. The green dots, to larger topology-optimized
cloaks, with $R=5.5\lambda$ and $h=15.5\lambda$. Both sets of data
overlap at ${P>3\cdot 10^{-2}\lambda}$. At lower pumping
strengths, however, the concurrence decays significantly with
decreasing $P$ in the small devices, while it grows towards
$C\approx 0.5$ in the large ones. Importantly, the data for the
latter resembles very much to those in Figure \ref{fig:2}(c), which
corresponded to smaller $d_{12}$.
We can identify the reduction of entanglement in small cloaks with finite-size effects, as the number of reflecting elements in the cloaks is not enough to prevent the occurrence of significant radiation loss.

\begin{figure}[t!]
\centering
\includegraphics[width=\linewidth]{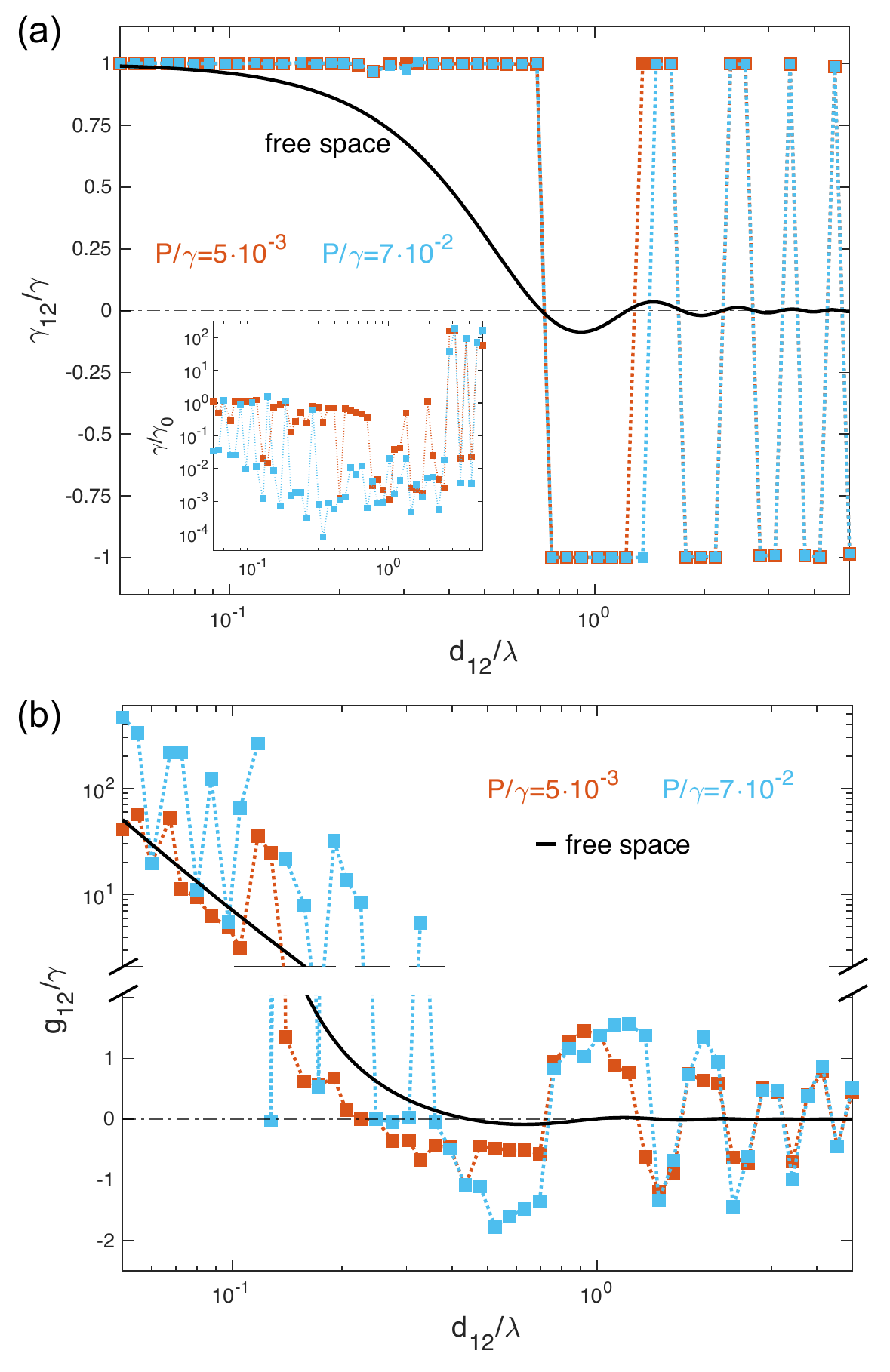}
\vspace{-0.6 cm} \caption{(a) Dissipative coupling strength
normalized to the QE decay rate as a function of the inter-emitter
distance for the two pumping rates in Figure \ref{fig:2}(b-c). The
inset plots the Purcell factor versus $d_{12}$ for the cloaks in
the main panel. (b) Normalized coherent coupling strength for the
same ID structures. The black solid lines in both panels
correspond to free-space magnitudes.} \label{fig:4}
\end{figure}

In order to shed light into the entanglement generation mechanism
taking place in the ID cloaks, we examine the dissipative,
$\gamma_{12}$, and coherent, $g_{12}$ coupling strengths that
results from the topology-optimization design. Figure \ref{fig:4}(a)
plots the former, normalized to the QE decay rate, as a function
of the inter-emitter distance and for the two pumping rates
considered in Figure \ref{fig:2}(b). For comparison, the same
magnitude evaluated in free space is rendered in black solid line.
We can observe that all the designs maximize the dissipative
coupling, so that $|\gamma_{12}|/\gamma=1$, while its sign follows
its free-standing counterpart. Note as well that the data for both
$P/\gamma$ overlap. These results demonstrate that the
topology-optimized dielectric structures generate entanglement
through the same dissipative-driven mechanism that occurs
naturally in metal-based plasmonic
nanostructures~\cite{GonzalezTudela2011,Hou2014,MartinCano2011}.
The inset of Figure \ref{fig:4}(a) displays the Purcell factor,
$F=\gamma/\gamma_0$, experienced by both QEs for the designs in
the main panel. It shows that the ID devices are capable of
implementing the maximum dissipative coupling condition for
extremely different QE decay rates. Remarkably, $\gamma/\gamma_0$
ranges 6 orders of magnitude in the cloaks. On the one hand,
$\gamma$ is reduced up to a factor $10^{-4}$ for
$d_{12}<3\lambda$. On the other hand, it becomes 100-fold enhanced
for larger inter-emitter distances, where the device efficiency
diminishes due to finite size effects. Figure \ref{fig:4}(b) displays
the coherent coupling in the cloaks, revealing that they introduce
only small deviations from free space. At small QE-QE distances,
$g_{12}\gg\gamma$, in the regime where the entanglement
enhancement by the cloaks, $C-C_0$, is moderate. On the contrary,
$|g_{12}|/\gamma\approx 1$ at longer $d_{12}$, where the coherent
coupling vanish in free space.

\begin{figure}[t!]
\centering
\includegraphics[width=\linewidth]{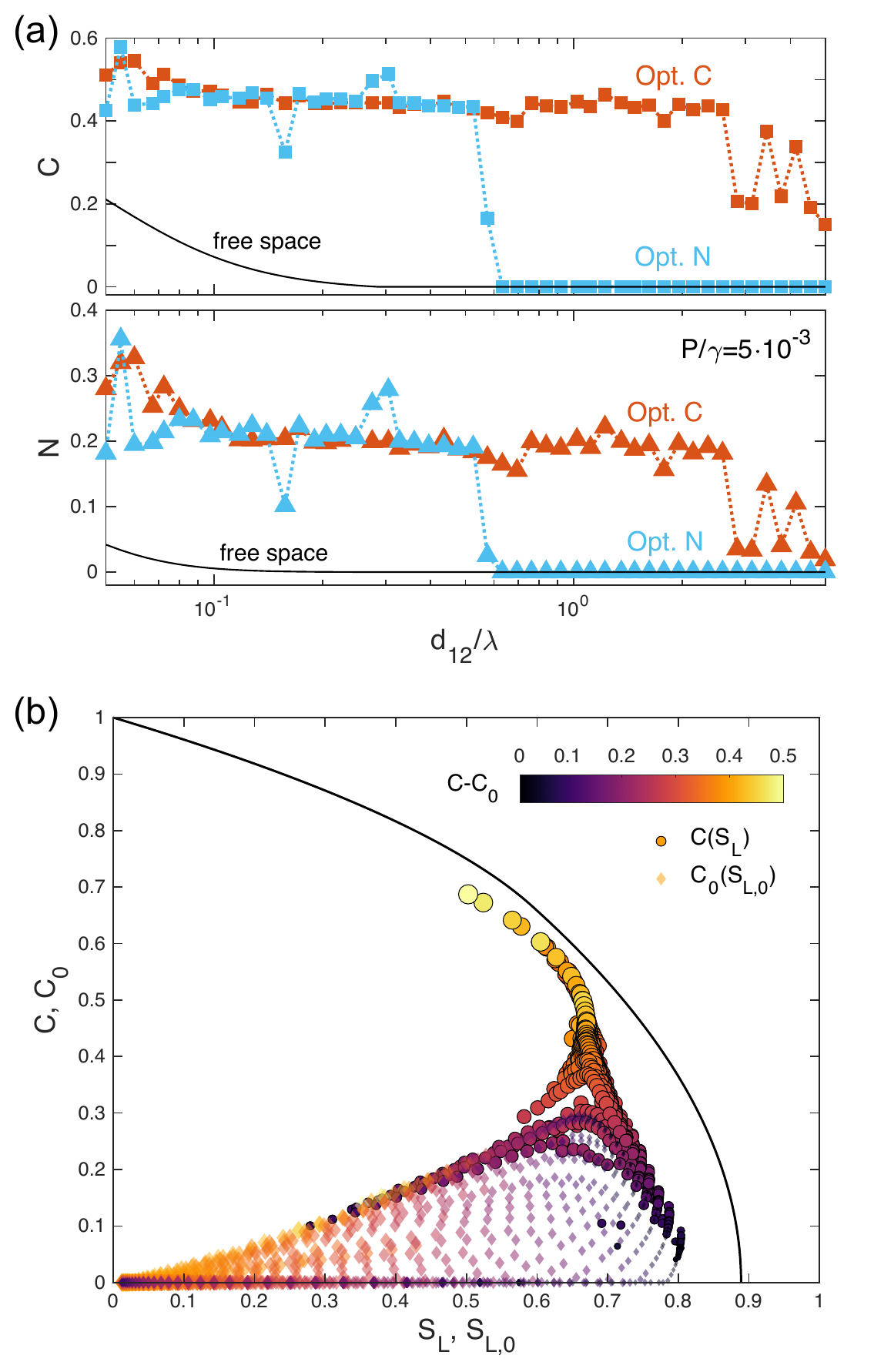}
\vspace{-0.4 cm} \caption{(a) Top: Concurrence versus
inter-emitter distance for dielectric structures obtained by
setting the concurrence (orange) and the negativity (blue) as the
optimization function ($P/\gamma=5\cdot 10^{-3}$). Bottom:
Negativity versus distance for the same designs as in the top
panel. In both cases, black solid lines represent the results for
free-standing QE pairs. (b) Concurrence versus linear entropy for
the ID cloaks in Figure \ref{fig:2}(a) (bright color circles),
together with the corresponding values for the QE pairs in free
space (faint color rhombuses). The colors code the concurrence
enhancement, $C-C_0$ for each system. The black solid line
corresponds to the maximally-entangled-mixed-states figure of
merit.} \label{fig:5}
\end{figure}

Up to here, we have considered only the Wootters concurrence as a
measure of entanglement. However, there exist multiple witnesses
that have been proposed for bipartite systems~\cite{Chru2014}.
Next, we take another, the (linear) negativity,
$N$~\cite{Vidal2002,Miranowicz2004}, to assess the suitability of
the Wootters concurrence, $C$, as the optimization function in our
ID approach. The negativity is defined in terms of the negative
eigenvalues of the partial transpose of the density matrix,
$\rho$. For our system, it has a very simple form as well,
\begin{equation}
N=N(\rho)={\rm
max}\{0,\sqrt{(\rho_{00}-\rho_{33})^2+4|\rho_{12}|^2}-(\rho_{00}+\rho_{33})\}.
\label{N}
\end{equation}
By simple inspection, we can conclude that, similarly to
$C$, entanglement formation ($N>0$) takes place under the
condition $|\rho_{12}|>\rho_{00}\rho_{33}$ in $N$. In
Figure \ref{fig:5}(a), we explore whether both equations also yield
the same dielectric structures when employed as the maximization
function in our topology-optimization algorithm. The top panel
plots the concurrence versus inter-emitter distance for cloaks
operating at $P/\gamma=5\cdot 10^{-3}$. Orange (blue) dots
correspond to the designs obtained for concurrence (negativity)
maximization, and the solid black lines plot $C_0$. We can observe
that for $d_{12}\lesssim\lambda/2$ both sets of devices yield the
same concurrence. The dielectric maps obtained from the
maximization of both magnitudes are the same in this regime. On
the contrary, at larger distances, the negativity-based algorithm
does not find the optimization path in the concurrence-based
procedure. This way, the outcome of the former is simply free
space. To shed light into this finding, the lower panel of
Figure \ref{fig:5}(a) plots the negativity for the same structures,
together with its free-space value, reproducing the same
behaviour. Note that $N\gg N_0$ at large distances only in the
designs obtained by setting $C$ as the optimization function.
These results manifest that, as expected, our gradient-based
topology-optimization approach is very sensitive not only to the
target function, but also to the initial conditions (always set to
free space in our study).

Finally, and once we have shown the dependence of the cloak
designs on the optimization function, we proceed to benchmark
their performance against
maximally-entangled-mixed-states~\cite{Ishizaka2000}. To do so, we
calculate first the linear entropy for all the QE-QE states
realized by the devices in Figure \ref{fig:2}(a). This way, we
establish their mixed/pure character. This quantity is defined in
terms of the trace of the density matrix squared~\cite{Vidal2002}.
In our case, it reads
$S_L=S_L(\rho)=\tfrac{4}{3}[1-\rho_{00}^2+\rho_{11}^2+\rho_{22}^2+\rho_{33}^2-2|\rho_{12}|^2]$,
being 0 for pure states, and 1 for maximally mixed ones.
Figure \ref{fig:5}(a) plots $C$ versus $S_L$ for our designs in
bright color circles. Faint colored rhombuses render $C_0$ as a
function of $S_{L0}$ for the same distance and pumping
configurations but in free space. In both sets of data, the colors
code the concurrence enhancement, $C-C_0$, for each value of
$d_{12}$ and $P$. This panel shows clearly that our ID structures
are most efficient when acting on QE-QE states that present a high
purity in free space ($S_{L,0}\lesssim 0.4$), while their impact
is lower in free-standing states that present a higher entropy.
This demonstrates that the designs enhance and generate
entanglement by increasing the mixed character of the emitter
states. 
Importantly, the black solid line in Figure \ref{fig:5}(b) presents the concurrence-entropy curve for maximally-entangled-mixed-states~\cite{Munro2001,delValle2011}. It
reveals clearly that the optimum cloaks approach greatly this
figure of merit, yielding the maximum entanglement attainable for
the linear entropy of the QE-QE state induced by the dielectric
structure.

\section{Conclusion}
To conclude, we have applied inverse design ideas to the problem
of bipartite entanglement generation under incoherent pumping
conditions. Through a topology-optimization algorithm that, acting
at the level of the electromagnetic Dyadic Green's function,
maximizes the Wootters concurrence, we have generated dielectric
cloaks hosting quantum emitter pairs for different distance and
pumping configurations. First, the entanglement enhancement
provided by these devices has been assessed, showing that they are
most efficient when operating on emitters that are completely
untangled in free space. Next, we have analyzed the permittivity
maps for these devices and explored the impact of finite-size
effects in their performance. We have also shown that they operate
by maximizing the dissipative coupling strength between the
emitters, even under extremely different Purcell enhancement
conditions. Finally, we have studied the dependence of the design
outcome on the entanglement witness used as the optimization
function, and have benchmarked our results against
maximally-entangled-mixed-states. We believe that our work
illustrates the power of inverse design as a tool to improve
quantum information resources based on nanophotonic platforms, and
opens the way towards the exploitation of similar approaches in
larger, more complex, quantum emitter networks.

\begin{acknowledgments}
This work has been sponsored by the Spanish MCIN/AEI/10.13039/50110001033 and by "ERDF A way of making Europe" through Grant Nos. RTI2018-099737-B-I00 and CEX2018-000805-M
(through the Mar\'ia de Maeztu program for Units of Excellence in R$\&$D). We also acknowledge funding from the 2020 CAM Synergy Project Y2020/TCS-6545 (NanoQuCo-CM). We thank Alejandro Gonz\'alez-Tudela and Michel Frising for their feedback and fruitful discussions.
\end{acknowledgments}

\bibliography{article}

\end{document}